\documentclass[twocolumn]{article}
\usepackage[letterpaper, portrait, margin=0.8in]{geometry}

\usepackage{graphicx}
\usepackage{abstract}
\usepackage{amsmath}
\usepackage{amssymb}
\usepackage{authblk}
\usepackage{multirow} 
\usepackage[T1]{fontenc}
\usepackage{natbib}
\begin{document}
\title{\bf Prospects for Wideband VLBI Correlation in the Cloud}
\author[1,2]{Ajay Gill}
\author[1]{Lindy Blackburn}
\author[1,3,4]{Arash Roshanineshat}
\author[4]{Chi-Kwan Chan}
\author[1]{Sheperd S. Doeleman}
\author[1]{Michael D. Johnson}
\author[1]{Alexander W. Raymond}
\author[1]{Jonathan Weintroub}

\affil[1]{Center for Astrophysics | Harvard \& Smithsonian, 60 Garden Street, Cambridge, MA 02138, USA}
\affil[2]{Department of Astronomy and Astrophysics, University of Toronto, 50 St. George Street, Toronto, ON  M5S 3H4, Canada}
\affil[3]{Department of Electrical and Computer Engineering, University of Arizona, 1230 E. Speedway Blvd., P.O. Box 210104, Tucson, AZ.}
\affil[4]{Department of Astronomy and Steward Observatory, University of Arizona, 933 North Cherry Avenue, Tucson, AZ 85721-0065}
\twocolumn[
  \maketitle            
  \begin{onecolabstract} 

\noindent This paper proposes a cloud architecture for the correlation of wide bandwidth VLBI data. Cloud correlation facilitates processing of entire experiments in parallel using flexibly allocated and practically unlimited compute resources. This approach offers a potential improvement over dedicated correlation clusters, which are constrained by a fixed number of installed processor nodes and playback units.  Additionally, cloud storage offers an alternative to maintaining a fleet of hard-disk drives that might be utilized intermittently. Here, we describe benchmarks of VLBI correlation using the \texttt{DiFX-2.5.2} software on the Google Cloud Platform to assess cloud-based correlation performance. In our analysis, the number of virtual CPUs (vCPUs) per Virtual Machine was varied to determine the optimum configuration of cloud resources. The number of stations was varied to determine the scaling of correlation time with VLBI arrays of different sizes. Data transfer rates from Google Cloud Storage to the Virtual Machines performing the correlation were also measured. Based on the results, we present an example cloud correlation configuration. Current cloud service and equipment pricing data is used to compile cost estimates allowing an approximate economic comparison between cloud and cluster processing. We note that the economic comparisons are based on cost figures which are a moving target, and are highly dependent on factors such as the utilization of cluster and media, which are a challenge to estimate. Our model suggests that shifting to the cloud is an alternative path for high data rate, low duty cycle wideband VLBI correlation that should continue to be explored. In the production phase of VLBI correlation, the cloud  has the potential to significantly reduce data processing times and allow the processing of more science experiments in a given year for the petabyte-scale data sets increasingly common in both astronomy and geodesy VLBI applications.  
  \end{onecolabstract}
  ]

\section{Introduction}
The technique of Very Long Baseline Interferometry (VLBI) links together many radio telescopes to make high angular resolution observations of distant astronomical sources, or in geodesy to precisely measure the shape and orientation of the Earth \citep{clark,moran,thompson}. The sensitivity for VLBI instruments is determined by the collecting area  \citep{Matthews_2017}, by the integration time, and by the recorded bandwidth. For a given antenna size and atmospheric coherence time, the sensitivity improves as $ B^{1/2}$, where $B$ is the total recorded bandwidth \citep{whitney}. The millimeter-wave VLBI array known as the Event Horizon Telescope (EHT), for example, has steadily increased its bandwidth from less than 0.5 to 16~GHz \citep{PaperII} or a data recording rate for 2-bit samples of 64~Gbps (gigabits-per-second).  Such a large bandwidth yields petabyte-scale recordings, which is an enormous data volume and therefore a challenge to correlate.

In typical VLBI arrays, each station simultaneously observes the astronomical target and records the data onto hard-disk drives that are then sent to a correlation facility, where hardware such as Field-Programmable Gate Arrays (FPGAs), Application-Specific Integrated Circuits (ASICs), Central Processing Units (CPUs), and/or Graphical Processing Units (GPUs) perform the correlation. Instead of transporting disk drives, some arrays transfer the data directly over network with a parallel file system from a station or a nearby staging area to a storage pool at the correlation facility. In this study, we primarily consider  disconnected-element interferometers with stations separated by long baselines ($\gtrsim$ 100 km), where the data are recorded on disk drives and transported to the correlation facility.  Doubling the received bandwidth to improve sensitivity means that the recording rate and the data volume also doubles for a given integration time.  Advances in high-speed recording made possible by industry-supported enhancements in analog-to-digital conversion \citep{Patel2014} and improvements in network data transfer mean that data volumes for modern VLBI arrays operating at wide bandwidths can be many petabytes per month. The trend to use more bits per sample to improve the signal-to-noise ratio and RFI robustness also translates to increasing  data rates and data volumes.    

To address the difficulty of processing these large data sets, we consider the possibility of using decentralized computational resources available over the Internet. These resources, referred to as the \textit{cloud}, enable the realization of Virtual Machines (VMs) consisting of multiple virtual Central Processing Units (vCPUs). Cloud-based VMs could process VLBI data in parallel at speeds commensurate with the increasing recording rates.  A benefit of the cloud approach to VLBI correlation is the ability to process more science experiments in a given year. A large dedicated supercluster, where the data is transferred in parallel over network from the stations or a nearby staging area provides similar improvements in data processing times, but it is not always possible for every research group to access such a supercluster.

There have been previous studies of distributed computing for VLBI. In early work, \citep{Huib} employed a precursor to the cloud called the {\it grid} in a project called {\it FABRIC}, where distributed computing was applied to the Merlin array observing in an e-VLBI mode with correlation in real time via the Internet. \citep{Weston} performed benchmarks for VLBI correlation using the Catalyst IT Ltd commercial cloud in New Zealand. They found that performing correlation on the cloud provides identical fringe outputs compared to cluster correlation, and they benchmarked a speed-up of a factor of $\sim$ 2.5, correlating data rather narrower in bandwidth compared to the present study. On the software side, \texttt{CorrelX} is a correlation software package currently under development to leverage the cloud for VLBI correlation\footnote{https://github.com/MITHaystack/CorrelX}. After experimenting with \texttt{CorrelX} and consulting with its authors, we elected instead to use the widely used \texttt{DiFX-2.5.2} VLBI software correlator \citep{Deller_2011} exclusively for this study.   

In this paper, we propose an architecture for shifting wideband VLBI correlation to the cloud. To assess the efficiency of cloud correlation, benchmarks across a parameter space of possible configurations were carried out on the Google Cloud Platform (GCP)\footnote{https://cloud.google.com}. Performance and scaling trends were characterized by varying the number of vCPUs per VM, the number of stations in the array, and testing different data transfer rates from Google Cloud Storage (GCS) to the VMs. Although the GCP was used for this study, the proposed cloud correlation architecture is generalizable to other major cloud platforms such as Amazon AWS\footnote{https://aws.amazon.com} and Jetstream\footnote{https://jetstream-cloud.org} \citep{JetStream,JetStream2}. 

\section{VLBI applications}
The two major applications of VLBI are astronomical research and geodesy.
\subsection{Astronomy}
The angular resolution of a VLBI radio interferometer is $\theta \simeq \lambda\,/{B_{\rm max}}$ in radians or $\theta \simeq 2\times 10^{5}\lambda\,/{B_{\rm max}}$ in arcseconds, where $\lambda$ is the observation wavelength, and ${B_{\rm max}}$ is the length of the maximum projected baseline. At a wavelength of 2~cm with intercontinental baselines of 10,000 km, an angular resolution of 0.4 mas can be achieved, whereas at the shorter wavelength of 1.3 mm with  baselines of 10,000 km, an angular resolution of 20~$\mu$as is possible \citep{PaperIV}, a resolution that is unparalleled in Earth-based diffraction limited observations.

The massive compute and storage resources provided by the cloud could handle the increased data rates and data volumes resulting from wider bandwidth VLBI observations. The scientific opportunities enabled by expanding the bandwidth of VLBI have been considered in various studies \citep{doel2004,walker2007,fish2013}. This includes microarcsecond astrometry of high mass star forming regions within the Milky Way \citep{reid2009}, stellar mass black holes \citep{mj2009,reid2011}, pulsars and tests of general relativity \citep{Deller_2007}, and satellites within the Local Group. Increasing VLBI bandwidths can also enhance the number of detectable targets, transients, and astrometric reference sources, since the number of detectable sources scales with bandwidth as \textit{N} $\propto$ \textit{B}$^{3/4}$ \citep{reid2014}. Improvements in imaging gravitational lens systems to study the properties of the relativistic jets in distant Active Galactic Nuclei \citep{koopmnans} and small scale structure near the cores of lensing galaxies can be expected \citep{mao,winn,zach}. More sensitive observations could also allow  constraining the properties of supernovae \citep{bartel} and gamma ray bursts \citep{taylor1998,bientenholz,Abbott:2017xzu} by resolving their afterglows as well as constraining the energetics of the explosions and the nature of the circumburst environment.   

At short wavelengths ($\lambda$ $\leq$ 1.3 mm), VLBI can resolve and image the emission near the event horizon of nearby supermassive black holes, presenting an opportunity to study general relativity in a strong gravity regime \citep{doeleman2009G}. Recent VLBI observations at 1.3~mm by the EHT have captured the first image of the 6.5 $\times$ 10$^{9}$ $M_{\odot}$ supermassive black hole at the center of the giant elliptical galaxy M87 \citep{PaperI}. Previous observations have detected $\sim$ 5 Schwarzschild radii structure at the base of the relativistic jet produced at the core of M87 \citep{shep2012,Akiyama:2015qta}. Observations at 1.3~mm have also resolved structures on the scale of a few Schwarzschild radii in the 4 $\times$ 10$^{6}$ $M_{\odot}$ black hole at the center of the Milky Way, Sagittarius A$^{*}$ \citep{shep2008,fish2011,john2015,Lu2018}.


\subsection{Geodesy}
Geodesy is the study of the geometric shape, rotation and orientation in space of Earth and its gravitational field. Since VLBI observations can be used to determine antenna positions with millimeter accuracy, the technique is also well suited for geodetic studies. VLBI observations for geodesy are performed by the International VLBI Service for Geodesy and Astrometry \citep{geodesy1}, together with the satellite range network and the Global Navigation Satellite System antenna network \citep{whitney}.

VLBI for geodesy is carried out by observing many extragalactic broadband radio sources that are uniformly distributed across the sky \citep{geodesy2}. Since fast slewing antennas are required to quickly observe different sources around the sky, geodetic VLBI antennas typically have smaller diameters. To compensate for the decrease in sensitivity due to the smaller antenna diameters, higher bandwidths are needed to reach an equivalent sensitivity. To observe a uniform source distribution, it is often necessary in geodesy to observe weaker sources, which in turn also requires higher bandwidths. 

Under the current VLBI Global Observing System (VGOS) operation, VLBI correlation is distributed to multiple computer clusters, where each cluster is constrained by the number of available playback units and the percentage of time allocated for geodesy\footnote{https://ivscc.gsfc.nasa.gov/about/org/components/co-list.html}. In contrast, shifting the correlation of geodetic VLBI to the cloud could provide greater user flexibility by leveraging the massive I/O and compute resources for a short period of time.

\section{Cloud correlation architecture}
\begin{figure*} [htb!]
\centering
  \includegraphics[width=13cm]{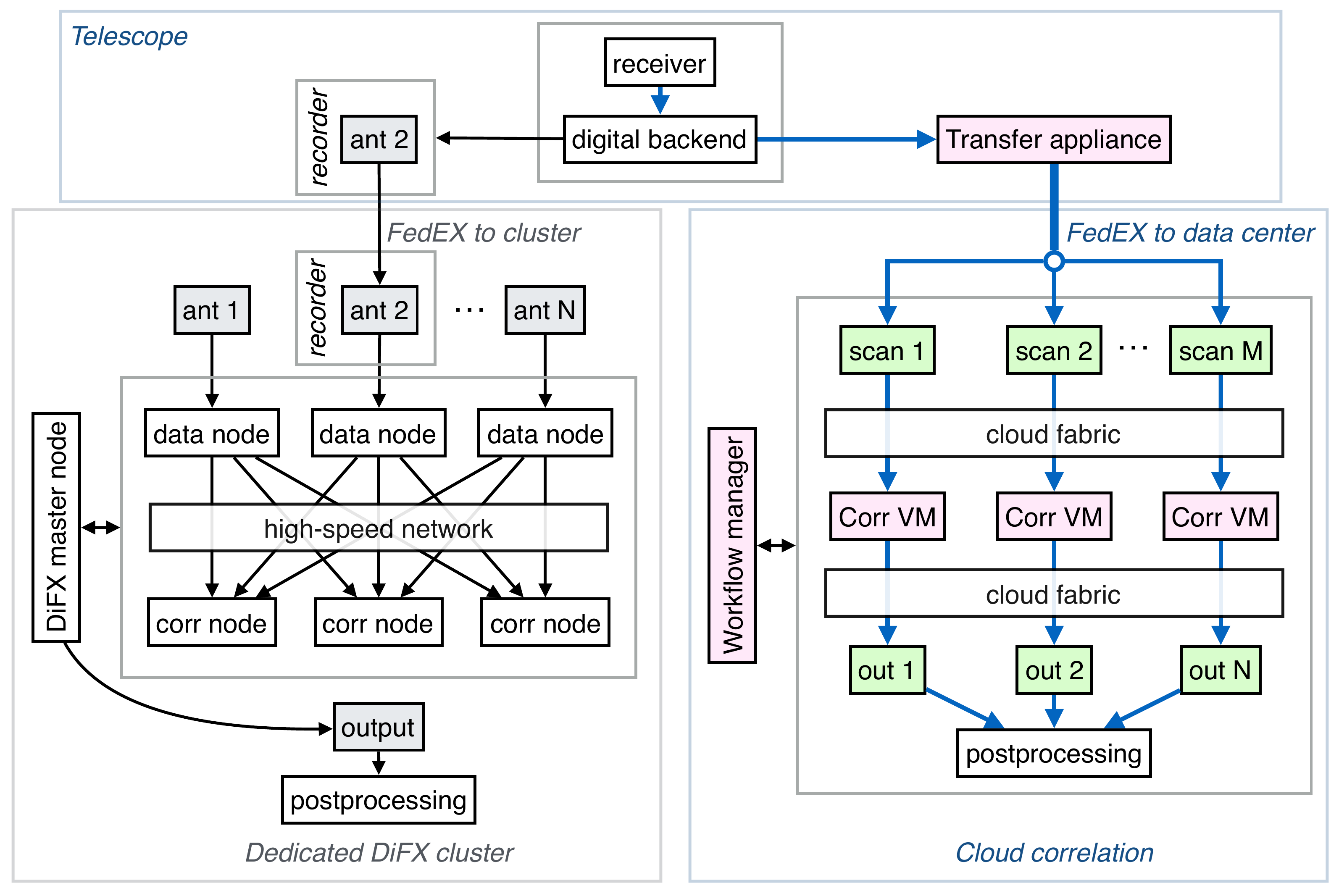}
  \caption{Comparison of standard cluster correlation architecture (left) with the proposed cloud correlation architecture (right). The proposed data flow is highlighted with blue arrows, where new data products are shaded green, and major new components in the development of the cloud pathway are shaded pink.  
  A typical cluster correlates the data streamed from recorders, where each cluster is limited by the number of available processor nodes and playback units as well as the percentage of cluster time allocated to different experiments. On the other hand, the cloud correlation architecture begins with the data separated by scans distributed across cloud storage staged using the Google Transfer Appliances and correlates the entire experiment in parallel using independent jobs. The ability to process data in parallel is equally true for a dedicated correlation facility consisting of a large supercluster that transfers data over network using a parallel file system, resulting in similar improvements in data processing times. Since not all research groups have access to a  supercluster, however, the cloud correlation framework provides greater accessibility and flexibility that enables researchers to leverage massive compute and storage resources for a short period of time. The cloud also allows international collaborators to monitor and process the data without the need to physically mount hard drives onto a local computing cluster.}
  \label{fig:hlevel}
\end{figure*}

We propose a cloud correlation architecture that leverages the parallel nature of the correlation workload and the massive parallel I/O provided by distributed cloud storage. Figure \ref{fig:hlevel} highlights the architectural differences between a typical VLBI dedicated cluster correlator and the proposed cloud correlation architecture. 

VLBI observations are normally separated into short integration scans, during which all antennas in the array simultaneously observe the same target source and typically record the data onto hard-disk recorders. An example of such hard-disk recorders are the Mark~6 recorders, which are disk-based data capture and record systems based on commercial off-the-shell (COTS) hardware and open-source software with a target sustained data capture rate of greater than 16~Gbps \citep{Mark6, whitney}. Google also provides similar bulk data recorders based on COTS hardware called Google Transfer Appliances (GTAs)\footnote{https://cloud.google.com/transfer-appliance/}.  We consider two possibilities for recording the data at the telescopes. The data could be directly recorded onto the GTAs. Alternatively, if the current GTA technology is not robust and reliable enough at present for recording at high altitudes and extreme temperatures, the data could initially be recorded onto the existing recorders and then transferred onto the GTAs. Recording onto the GTAs was not tested in this study, but it is not unreasonable to expect that the next generation GTAs could function with a similar capacity, speed, and reliability as the existing VLBI recorders. 

Once on the GTAs, the data are shipped to a data center, where they are staged onto the Google Cloud Storage (GCS). Multiple VMs can now access the raw data independently, enabling the correlation of many scans to proceed in parallel. The parallel processing of data is equally true for a dedicated correlation facility that uses a large parallel file system as a staging area.

The Workflow Manager in Figure~\ref{fig:hlevel} represents a suite of potential management software that can orchestrate and monitor the correlation and post-processing of data on the cloud. This includes \textit{pipeline generation scripts} that initialize a specific correlation or post-processing task in terms of a sequence of processes and dependencies, and a \textit{job scheduler} that executes the pipeline and dynamically allocates VM instances as needed. Although a Workflow Manager was not used in this study, Kubernetes\footnote{https://kubernetes.io} is an example of an open-source, industry-wide system that could be used as a Workflow Manager platform for cloud correlation.  For the benchmarks in this study, Docker\footnote{https://docs.docker.com} containers were used to create an ad-hoc scheme that packaged correlation and post-processing software along with their dependencies for deployment to the GCP. 

\section{Benchmark correlation setup}

We performed benchmarks based on two test data sets derived from a single recording made on an EHT backend system \citep{Vertat2015}. This recording was captured at a 2~GHz bandwidth and 2-bits per sample, which when critically sampled corresponds to an 8~Gbps data rate per polarization. Two orthogonal polarizations were recorded, and the data processing produces full polarization correlation products. The two benchmark data sets consist of copies made from the single recording to create synthetic arrays, one of 10 virtual stations, and the other of 20 virtual stations.  Thus, all the virtual stations in the test data sets have an effective 2~GHz bandwidth, dual-polarization, 2-bit sampling, and 20-second duration subscans, and all stations are copies of the same data. To ensure realistic processing and geometric calculations, each station was given an Earth-centered coordinate corresponding to a current EHT VLBI site location.  For each 20-second duration subscan, the 10-station data set has a total data volume of 400 GB; the 20-station data set is 800 GB. The 20-second duration subscans were used to easily accommodate the data size into the memory of the VMs. Since VLBI observation integrations are typically of longer duration, the 20-second benchmark results can be scaled linearly to longer scans.

The benchmarks carried out in this work could have been performed on either synthetically generated pure noise or recordings of noise in a laboratory setting, but it was convenient to use EHT recordings to verify the cluster versus cloud agreement for data taken under nominal astronomical observing conditions. For verification, benchmarks using \texttt{DiFX-2.5.2} were also performed on the cluster at the Smithsonian Astrophysical Observatory consisting of eight 10-core Intel Xeon E5-2640 v4 processors and one 16-core Intel Xeon E5-2620 v4 processor. The results of running the correlation on the GCP and on the cluster on identical datasets provided excellent agreement.

The GCP contains various types of VM instances based on the number of vCPUs and memory. For the \texttt{n1} series of machine types, a vCPU is implemented as a single hardware hyper-thread on one of the available CPU platforms. For the benchmarks, the \texttt{n1-highmem-16}, \texttt{n1-highmem-32}, \texttt{n1-highmem-64}, \texttt{n1-highmem-96}, and the \texttt{n1-megamem-96} VMs were used.  The specifications and price per hour for these VMs\footnote{https://cloud.google.com/compute/pricing\#machinetype} are given in Table~\ref{VMs}. VMs with large memory were used to easily accommodate the data volume of the test data sets in memory. The CPU platforms used were the Intel Xeon E5 (Sandy Bridge), Intel Xeon E5 v2 (Ivy Bridge), Intel Xeon E5 v3 (Haswell), and Intel Xeon E5 v4 (Broadwell E5) for the \texttt{highmem} VMs up to \texttt{n1-highmem-64}, and either of the four (Platinum, Gold, Silver, and Bronze) types of the Intel Xeon Scalable Processor (Skylake) for the \texttt{n1-highmem-96} and \texttt{n1-megamem-96} VMs\footnote{https://cloud.google.com/compute/docs/cpu-platforms}.

The correlation was performed on the \texttt{DiFX-2.5.2} software correlator. All available vCPUs on a VM were used as the \texttt{DiFX} core data processing nodes, even the ones on which the Datastream or the FXManager processes are run. One thread per core data processing node was used. The correlation parameters were set up with a Fast Fourier Transform (FFT) spectral resolution of 0.015625 MHz, an FFT window size of 64 $\mu$s, an FFT size of 262144, an integration time of 0.512 seconds, and an output spectral resolution of 0.5 MHz.

\section{Results}
The benchmarks were run varying the number of vCPUs and number of stations, and data transfer rates were also investigated.
\begin{table*}[htb!]
\centering
\begin{tabular}{|c|c|c|c|c|}
\hline
\textbf{Machine type} & \textbf{Number of vCPUs} & \textbf{Memory} & \textbf{Price/hr (USD)} & \textbf{Preemptible price/hr (USD)} \\ \hline
\texttt{n1-highmem-16} & 16 & 104 GB & \$0.94 & \$0.20 \\ \hline
\texttt{n1-highmem-32} & 32 & 208 GB & \$1.89 & \$0.40 \\ \hline
\texttt{n1-highmem-64} & 64 & 416 GB & \$3.79 & \$0.80 \\ \hline
\texttt{n1-highmem-96} & 96 & 624 GB & \$5.68 & \$1.20 \\ \hline
\texttt{n1-megamem-96} & 96 & 1433.6 GB & \$10.67 & \$2.26\\ \hline                    
\end{tabular}
\caption{Specifications and price per hour of the Virtual Machines on the Google Cloud Platform used in the benchmarks. The quantities are accurate as of the time of writing.}
\label{VMs}
\end{table*}

\subsection{Number of vCPUs}
To estimate the optimal number of vCPUs per VM to use for computation, we used the \texttt{n1-highmem-16}, \texttt{n1-highmem-32}, \texttt{n1-highmem-64}, and \texttt{n1-highmem-96} VMs, with the number of vCPUs varied from 16, 32, 64, and 96, respectively, as well as the number of stations varied from 2 to 10 stations. The 96-vCPU maximum was set by the ready availability of VMs up to that size on the GCP.

\begin{figure*} [htb!]
\centering
  \includegraphics[width=17cm]{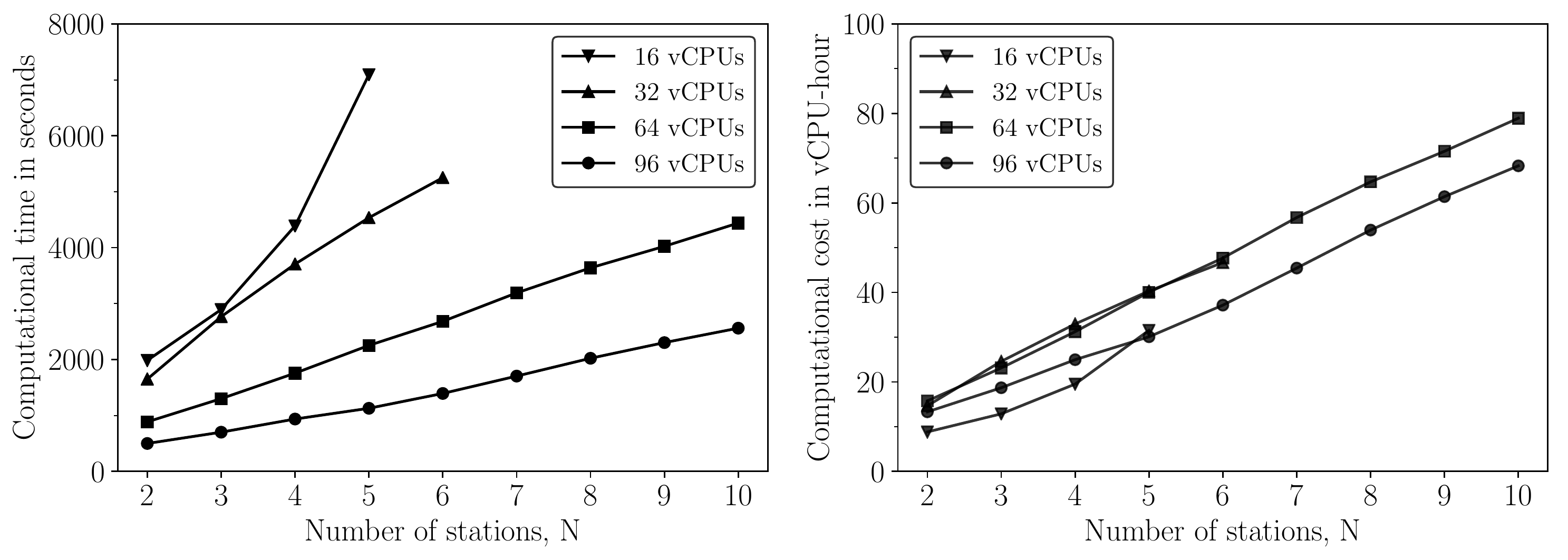}
  \caption{Computational time in seconds (left) and computational cost in vCPU-hour (right) with number of stations on the \texttt{n1-highmem-16}, \texttt{n1-highmem-32}, \texttt{n1-highmem-64}, and \texttt{n1-highmem-96} VMs. The results suggest that using 96 vCPUs per VM is the most efficient for data processing.}
  \label{fig:cores2}
\end{figure*}

\begin{figure} [htb!]
\centering
  \includegraphics[width=\columnwidth]{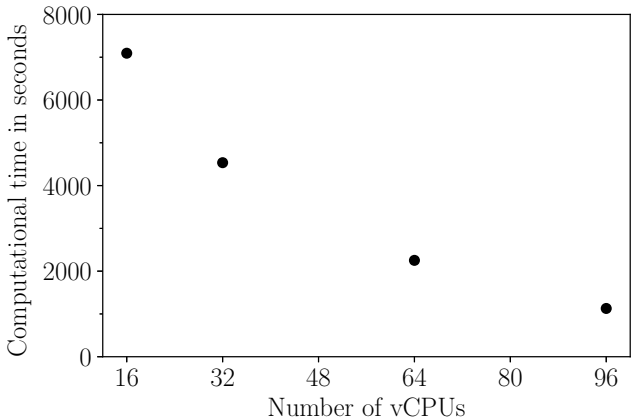}
  \caption{Computational time in seconds with number of vCPUs for a 5-station dataset on \texttt{n1-highmem-16, n1-highmem-32, n1-highmem-64, and n1-highmem-96} vCPU VMs. The result suggests an approximately inverse relationship between computational time and number of vCPUs.}
  \label{fig:cores}
\end{figure}

The results of computational time and cost with number of stations for different VM vCPUs are shown in Figure~\ref{fig:cores2}. The 16 and 32-vCPU benchmarks were only done up to 5 and 6 stations, respectively, since the computation for even more stations in those cases would have been too time-consuming. Figure~\ref{fig:cores} shows the computational time with the number of vCPUs for a 5-station dataset, suggesting an approximately inverse relationship between computational time and the number of vCPUs. 

To estimate the uncertainty in the measurements, the 96-vCPU, 10-station measurement was performed twice. The difference in correlation time was less than 1\%. For the 10-station test data set, these results suggest that using 96 vCPUs per VM is the most efficient for data processing.  

\subsection{Number of stations}
\begin{figure*} [htb!]
\centering
  \includegraphics[width=14cm]{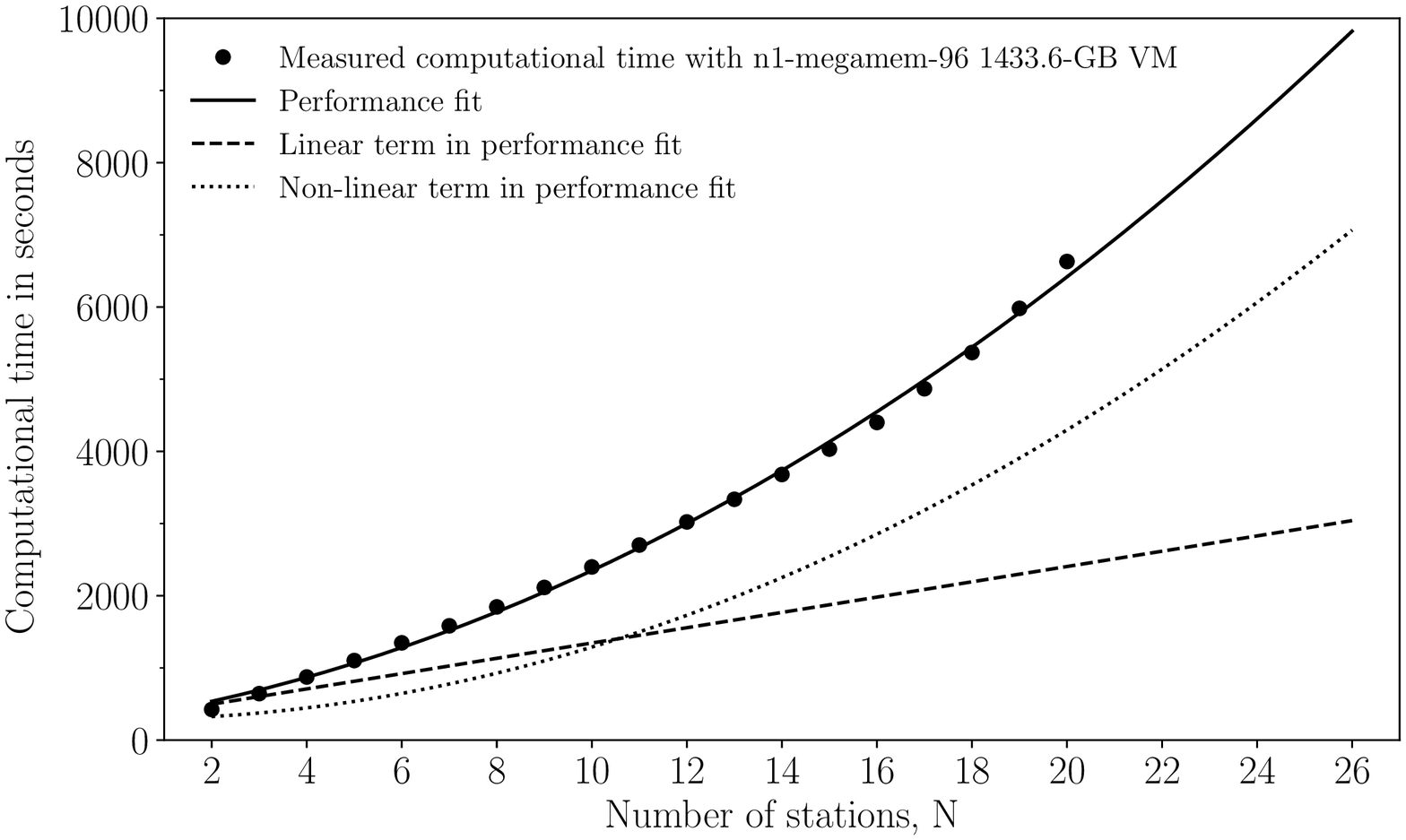}
  \caption{Correlation time with number of stations of a 20-second, 20-station, dual-polarization data set with a recorded bandwidth of 2 GHz per polarization and a data volume of 800 GB. The correlation was performed on the \texttt{n1-megamem-96 1433.6 GB} VM launched at the \texttt{us-west1} data center in Oregon. The linear term is expected due to the FFT per station, and the quadratic term is expected due to the channel-by-channel cross multiplications per baseline following the FFT performed by the \texttt{DiFX-2.5.2} software. The crossover point between the linear and non-linear terms is N $\sim$ 11.}
  \label{fig:compare}
\end{figure*}

To study the scaling of correlation time and thus computational cost with the number of stations, N, we used the 96-vCPU, \texttt{n1-megamem-96, 1433.6 GB} VM at the \texttt{us-west1} data center in Oregon. We varied the number of stations from 2 to 20. The results are shown in Figure \ref{fig:compare}. To estimate the uncertainty in the measurements, the 20-station measurement was performed twice, and the difference in correlation time between the two measurements was less than $0.3$\%. The measured computational time in seconds as a function of the number of stations was fit with a quadratic with a R$^{2}$ of 0.998.


\begin{equation}
t \simeq 1003\bigg(\frac{\rm N}{10}\bigg)^{2} + 1060\bigg(\frac{\rm N}{10}\bigg) + 285.3
\end{equation}

 The processing time is expected to exhibit a linear dependence on station number because many processing steps in \texttt{DiFX-2.5.2} are performed once per station.  A quadratic term in station number is also expected because the channel-by-channel cross-multiplications that follow the FFT are done for each baseline. The non-linear term begins to dominate at large N with a crossover point at N $\sim$ 11. The number of full-polarization correlation products is 4N$^{2}$, including auto-correlations. 

The deviation from purely quadratic dependence at large N could potentially be due to each Datastream and the FXManager processes taking a significant fraction of one vCPU's capacity, effectively limiting the number of total vCPUs performing the core processing. The set up time, which is the time from when \texttt{DiFX-2.5.2} is instantiated until the first bits begin to get correlated, and the tear down time, which is the time from when the last bits are correlated to when all the processes are terminated, could be substantial for the relatively short 20-second subscans used for the benchmarks. This effect is almost independent of the number of stations. The slowdown at large N could also be due to the extra memory required because of CPU cache misses described in \citep{Deller_2011}. The performance at large N could potentially be improved by further optimizing the \texttt{DiFX-2.5.2} correlation parameters.

\subsection{Data transfer rate}
\begin{figure*} [htb!]
\centering
  \includegraphics[width=16cm]{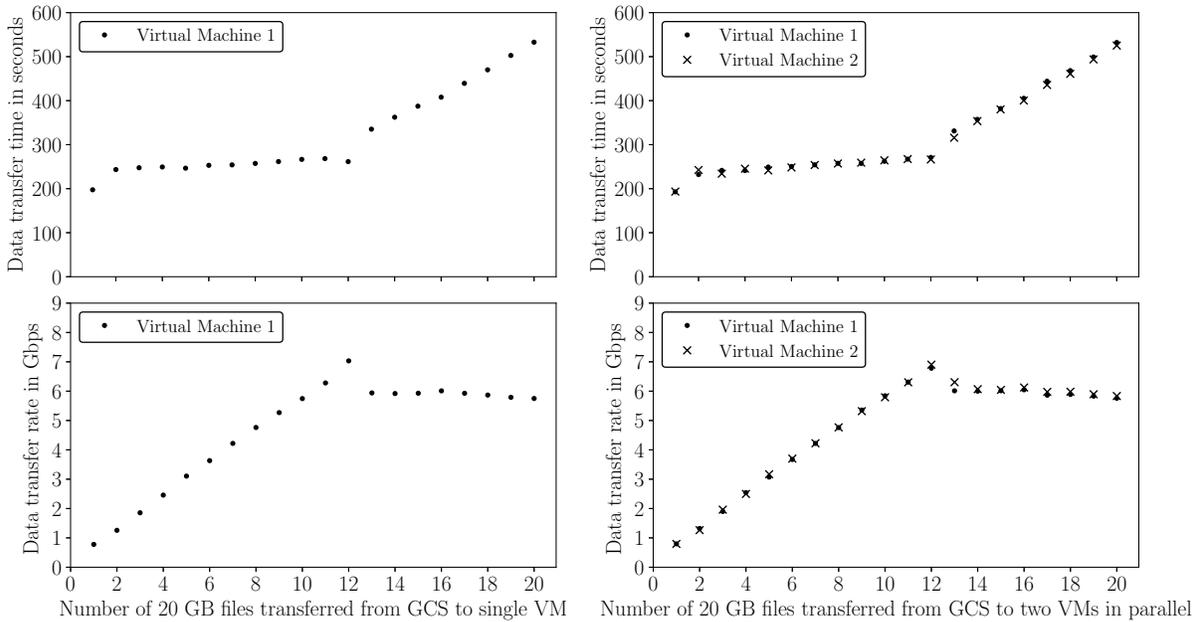}
  \caption{Data transfer rate and transfer time for transferring of VLBI Data Interchange Format (VDIF) files with a total size of 400 GB from the Google Cloud Storage to a single \texttt{n1-highmem-96, 624 GB} VM (left) and two of the same VMs in parallel. The aggregate data transfer rate increases as the number of files increases until about 12 files (240 GB), after which it saturates at $\sim$ 6 Gbps. The data transfer rate is not affected  when transferring the same data to two VMs in parallel.}
  \label{fig:datatransfer}
\end{figure*}

We transferred 10-station VLBI data consisting of a single 20-second subscan as 20 VLBI Data Interchange Format (VDIF) files each with a size of 20 GB making a total size of 400 GB, from the GCS to \texttt{n1-highmem-96, 624 GB} VM to measure the data transfer rate. We then transferred the same data to two different VMs in parallel to study its effect on the data transfer rate. The results are shown in Figure \ref{fig:datatransfer}.

First, the aggregate data transfer rate to a single VM increases as the number of files increases until about 12 files (240 GB), after which it saturates at $\sim$ 6 Gbps. Second, the data transfer rate is not affected when transferring the 10-station data to two VMs in parallel, as shown in the right panel of Figure \ref{fig:datatransfer}. These results suggest that it is not unreasonable to expect that VLBI data separated by scans can be independently transferred from the GCS to several VMs without a decrease in the data transfer rate, since the cloud fabric is designed to support independent resource scaling while simultaneously catering a large collection of users at any given time. 

\subsection{Example correlation configuration} \label{lowlevel}
\begin{figure*} [htb!]
\centering
  \includegraphics[width=\textwidth,height=\textheight,keepaspectratio=true]{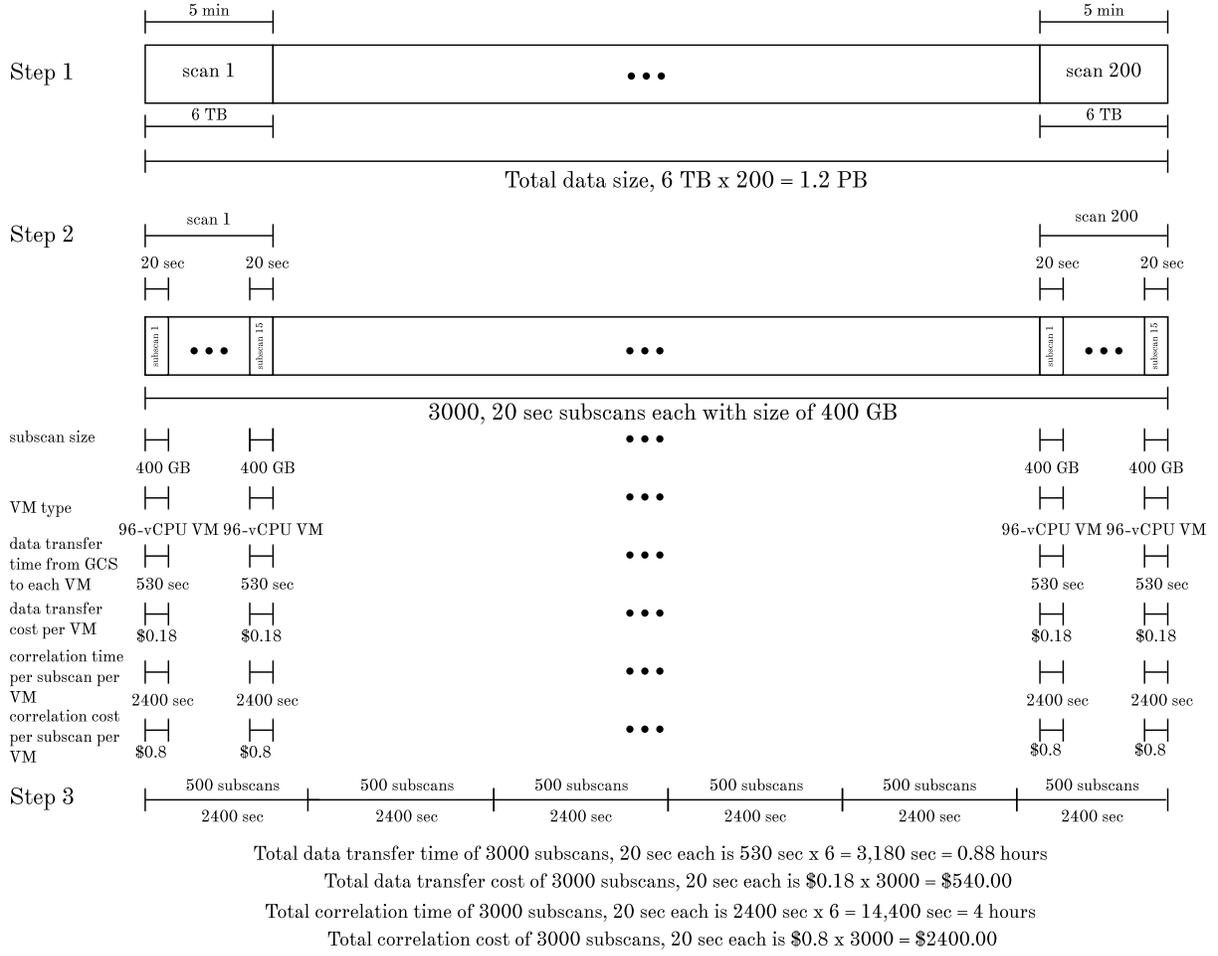}
  \caption{An example cloud correlation configuration of a 10-station, dual-polarization VLBI data set with 2 GHz bandwidth per polarization, a total size of 1.2 PB, and a total observation time of $\sim$ 16.7 hours. A total of 200, 5-minute scans are subdivided into 3000, 20-second subscans each with a size of 400 GB. Each 5-minute scan and 20-second subscan already consists of data from all 10 stations. Each 20-second subscan is transferred from the Google Cloud Storage to 3000 individual \texttt{n1-highmem-96} VMs. Each individual VM correlates a 20-second subscan in $\sim$ 2400 seconds, which corresponds to a total correlation time of $\sim$ 4 hours ($\sim$ 25\% of total exposure time) if processed with 6 consecutive instances of using 500 VMs in parallel at a time.  For this example correlation configuration, the total data transfer cost is \$540, and the total computational cost is \$2400.}
  \label{fig:flowchart}
\end{figure*}

Based on the benchmark results, we describe an example cloud correlation configuration in Figure \ref{fig:flowchart}. In all sections that follow, we consider the processing of a 10-station dual-polarization VLBI observation consisting of a total of 200, 5-minute individual scans, constituting a total observation time of $\sim$ 16.7 hours and a total data size of 1.2 PB recorded at a bandwidth of 2 GHz per polarization. Each 5-minute scan or 20-second subscan referred to in Figure \ref{fig:flowchart} already contains data from all 10 stations. 

The first step consists of staging the data from the GTAs onto the GCS. Each 5-minute scan from all 10 stations consists of 6 TB of data volume, which is too large to transfer onto any individual VM.

The second step involves pre-processing. The correlation parameters are initially set up, after which the data are split into smaller time segments or multiple frequency channels or transformed into the frequency domain for faster correlation. For instance, each 5-minute scan could be divided into smaller time segments that are small enough to load onto individual VMs. The 5-minute scans can be split into 15, 20-second subscans, each with a size of 400 GB, making a total of 3000 subscans for the entire dataset. These subscans can then be transferred from the GCS to 3000 independent \texttt{n1-highmem-96, 624 GB} VMs in parallel. The data transfer benchmark results in Figure~\ref{fig:datatransfer} suggest that it is not unreasonable to expect that the parallel data transfer rate will not be affected when transferring to multiple VMs with the assumption that bottlenecks are not encountered. The transfer of 3000 scans should take $\sim$ 530 seconds per VM. Future optimization could allow the transfer of data as it being processed, as further discussed in Section \ref{data_transfer_to_VM}.

The third step consists of the data processing. To make the computation more manageable, the correlation could be performed using six consecutive instances of simultaneously using 500 subscans or 500 VMs in parallel. From Figure \ref{fig:cores2}, the expected \texttt{DiFX-2.5.2} correlation time of a 10-station, 20-second subscan with a size of 400 GB subscan is $\sim$ 2400 seconds per VM. The total expected correlation time for all 3000 subscans is $\sim$ 4 hours ($\sim$ 25\% of total exposure time).

The total time for which the data is stored in the cloud will be dominated by the time required to set up the optimal correlation parameters in the pre-processing step and the scrutinization of the correlated products afterwards. It is sometimes necessary to perform multiple trial correlations to determine the optimal correlation parameters, troubleshoot bugs in the software correlator, and fix errors in the recorded data. An advantage of cloud correlation is the faster turn-around time on running trial correlations, which could make the process of determining the optimal correlation parameters much more efficient for the correlator operator. The additional time for setting up the correlation parameters and scrutinizing the correlated data products will generally be greater for part-time arrays compared to dedicated full-time arrays like the Very Long Baseline Array and geodetic arrays such as VGOS.  

\section{Cloud cost estimate}
Based on the benchmark results, the costs for processing the example 10-station observation on the cloud can be estimated. The various components that contribute to the overall cost including the rental of recording units, shipping, data storage, data transfer, and computation are discussed in this section.

\subsection{Recording and shipping}
Here, we consider recording directly onto to the GTAs at the telescopes with the assumption that either the current or the next generation GTAs are reliable enough to record data at high altitudes. The alternative is to first record the data onto existing VLBI recorders and then transfer the data to the GTAs. Recording the 1.2 PB data would require twelve 100 TB GTAs, each with a usage fee of \$300 with 10 free days of on-site usage. The total cost would be 12 $\times$ \$300 $=$ \$3,600.  The two-way shipping cost of the GTAs for a 16 Gbps observation is estimated at \$100 per site, or  $\sim$ \$1,000.

\subsection{Storage}
Data storage dominates the cost of the cloud correlation architecture. On the GCS, there are four types of storage classes\footnote{https://cloud.google.com/storage/docs/storage-classes\#regional}: regional, multi-regional, nearline, and coldline. 

Regional storage consists of data storage in only one specific region with no redundancy. If storage of data only in a single region with no redundancy is acceptable for the experiment, which can be expected to be the case for most VLBI experiments, then regional storage could be used. Regional storage costs \$0.02 per GB per month. The storage cost model on the GCS is not monthly based, and per-day pricing is also available.  

Multi-regional storage is designed for frequently accessed data, is geo-redundant (stored in at least two different geographic locations), and ensures maximum data availability, but it is more expensive than regional storage. It costs \$0.026 per GB per month.  

Nearline and coldline storage are low-cost storage options for infrequently accessed data, once a month for nearline and once a year for coldline storage. The cost is \$0.01 per GB per month for nearline storage and \$0.007 per GB per month for coldline storage. However, both nearline and coldline storage have a retrieval cost of \$0.01 per GB and \$0.05 per GB, respectively.

Since many radio telescopes are located at remote places such as the South Pole and on top of mountains, it is not unreasonable to expect delays in acquiring the GTAs from some stations in the VLBI array. Therefore, the cheaper nearline or coldline storage could be used for longer term storage until the GTAs from all stations have been acquired, with the caveat of retrieval costs. We consider the storage of 1.2 PB in regional storage for one month for a cost estimate of \$25,166. This represents a best-case lower bound, and assumes no major delays from telescopes at remote locations. Where such delays exist, costs for GTA rental and cloud storage may be substantially higher, and nearline or coldline storage needs to be specified at a minimum.

\subsection{Data transfer to Virtual Machines}\label{data_transfer_to_VM}
In our benchmark operational model, correlation of data does not begin until raw data from all antennas has been transferred in full to the VMs.
The data transfer of 3000 scans to the VMs in parallel should take $\sim$ 530 seconds per VM, which would cost \$0.18 per VM with a total cost of \$0.18 $\times$ 3000 $=$ \$540 for all 3000 subscans, considering a preemptible price of \$1.20 per hour for the \texttt{n1-highmem-96, 624 GB} VM.
With relatively minor changes to the underlying software and data model, however, it would be possible instead to stream data as it is being processed and remove any separate cost for the data transfer time. The \texttt{DiFX-2.5.2} software correlator supports a streaming mode in which the data can be accepted via a network socket. Since data transfer bandwidth exceeds the rate of correlation, this would not impact correlation run-time. Because we did not use a streaming data model for the model, however, we include the costs of transfer time in Table \ref{totalcost}.

\begin{table*} [htb!]
\centering
\begin{tabular}{|l|c|c|c|c|c|c|}
\hline

\multirow{2}{*}{Virtual Machine type} &
\multicolumn{1}{c|}{Record \& ship}                  & \multicolumn{1}{c|}{Storage} & \multicolumn{1}{c|}{Transfer} & \multicolumn{1}{c|}{Computation} & \multicolumn{1}{c|} {\textbf{Total}} &
\multicolumn{1}{c|} {\textbf{Total}} \\ 

&
\multicolumn{1}{c|}{(16 Gbps)}                  & \multicolumn{1}{c|}{(16 Gbps)} & \multicolumn{1}{c|}{(16 Gbps)} & \multicolumn{1}{c|}{(16 Gbps)} & \multicolumn{1}{c|} {\textbf{(16 Gbps)}}  &
\multicolumn{1}{c|} {\textbf{(64 Gbps)}} \\

\hline
Preemptible VMs     & \$4,600     & \$25,166     & \$540      & \$2,400  & \textbf{\$32,706} &
 \textbf{\$130,824} \\ \hline
Non-preemptible VMs & \$4,600     & \$25,166      & \$2,510      & \$11,370 & \textbf{\$43,646} & \textbf{\$174,584} \\ \hline
\end{tabular}

\caption{Cloud cost estimate for the \texttt{DiFX-2.5.2} software correlation of a 10-station, dual-polarization VLBI dataset recorded at 16 Gbps with 2 GHz bandwidth per polarization and an observation time of $\sim$ 16.7 hours on the Google Cloud Platform. The data storage cost is for the storage of 1.2 PB of data for one month in regional storage in South Carolina (\texttt{us-east1)}. A typical EHT VLBI experiment is recorded at 64 Gbps, which quadruples the 16 Gbps costs, as shown in the rightmost column and is used as the comparable to cluster cost in Section \ref{cluster_cost_sect}.}
\label{totalcost}
\end{table*}

\subsection{Computation}
The \texttt{DiFX-2.5.2} correlation time of a 400~GB subscan is expected to take $\sim$ 2400 seconds, which would cost \$0.80 per VM with a total cost of \$0.80 $\times$ 3000 $=$ \$2,400 for all 3000 subscans, considering a preemptible price of \$1.20 per hour for the \texttt{n1-highmem-96, 624 GB} VM. Here, we do not consider some up front computational cost of running trial correlations in the estimate.

Preemptible VMs are affordable, short-lived VMs, which last up to 24 hours. The disadvantage of using a preemptible VM is that it could be terminated at any time by the GCP if other higher priority compute instances require that particular VM. However, since the results suggest that the correlation time is $\sim$ 4 hours, and the benchmarks were performed using preemptible VMs, we expect that preemptible VMs would be appropriate for cloud correlation.  Non-preemptible VMs could also be used if necessary.
The correlation cost for a 400~GB subscan would be \$3.79 per VM with a total cost of \$3.79 $\times$ 3000 $=$ \$11,370 for all 3000 subscans, considering a non-preemptible price of \$5.68 per hour for the \texttt{n1-highmem-96, 624 GB} VM. 

We estimate the total cost for the cloud correlation of the full example observation in Table~\ref{totalcost}. For the 16 Gbps recorded benchmark data set, using preemptible VMs, the total cost is \$32,706, whereas using non-preemptible VMs, the total cost is \$43,646.  A typical EHT experiment, however, is recorded at 64 Gbps, which quadruples the 16 Gbps costs to \$130,824 and \$174,584 for preemptible and non-preemptible VMs, respectively.

\section{Cluster cost estimate} \label{cluster_cost_sect}

To provide an economic comparison between the cloud and a physical cluster, we consider the cost of a 1000-core, CPU-based cluster. Such a cluster is capable of correlating a 16 Gbps, 10-station VLBI experiment in a time period four times longer than the corresponding observation cumulative scan time.

\subsection{Cluster hardware} 
Equipment for this hypothetical cluster includes: 25 high-performance rack mount Linux machines, each with a dual-node with each node based on an Intel Xeon Processor with 20 cores; fast Ethernet switches and cables to tie the nodes together; 10 Mark6 playback machines each capable of delivering data at a rate of 16 Gbps; and physical infrastructure such as racking, uninterruptable power supplies (UPS) and network and power distribution cables. We have summed quoted costs for this equipment and arrive at an estimated capital investment of about \$500,000.  Assuming a five-year lifetime for the cluster, the cost is amortized linearly at \$100,000 per year.  

\subsection{Media}
In the cluster case, GTAs are not rented, so 1.2 PB of hard disk drive space needs to be purchased. We round up slightly given that the Mark6 disks have to be installed in 8-pack disk modules. At the time of writing, 10 terabyte enterprise quality Helium disk drives are selling for \$330, so this scales to  \$42,240.  Allowing for 8-pack disk modules for the Mark6, 16 modules at \$500 each, the media cost is estimated at \$50,000.  This is likewise linearly amortized over five years or \$10,000 per year.  

\begin{table*} [htb!]
\centering
\begin{tabular}{|l|r|r|r|r|r|r|r|}
\hline
\multirow{2}{*}{Cost type} & \multicolumn{1}{c|}{Cluster} & \multicolumn{1}{c|}{Media} & \multicolumn{1}{c|}{Cluster} & \multicolumn{1}{c|}{Media} & \multicolumn{1}{c|}{IT +} & \multicolumn{1}{c|}{\textbf{Total}} & \multicolumn{1}{c|}{\textbf{Total}}\\ 

\multicolumn{1}{|l}{} & \multicolumn{1}{|c|}{(16 Gbps)} & \multicolumn{1}{c}{(16 Gbps)} & \multicolumn{1}{|c}{(64 Gbps)} & \multicolumn{1}{|c}{(64 Gbps)} & \multicolumn{1}{|c|}{Elect.} & \textbf{(16 Gbps)} & \multicolumn{1}{c|}{\textbf{(64 Gbps)}} \\ 
\hline

{Capital}    & \multirow{2}{*}{\$500,000}     & \multirow{2}{*}{\$50,000}      & \multirow{2}{*}{\$500,000}      & \multirow{2}{*}{\$200,000}  & \multirow{2}{*}{\$700,000} & \multirow{2}{*}{\textbf{\$1,250,000}} & \multirow{2}{*}{\textbf{\$1,400,000}} \\

(no prorate)    & &  &  & &  &  &  \\ \hline

{Annual}   &  \multirow{2}{*}{\$100,000} &  \multirow{2}{*}{\$10,000}      &  \multirow{2}{*}{\$100,000} &  \multirow{2}{*}{\$40,000}  &  \multirow{2}{*}{\$140,000} &   \multirow{2}{*}{\textbf{\$250,000}} &  \multirow{2}{*}{\textbf{\$280,000}} \\ 
(no prorate)    & & & & & & & \\ \hline 

Annual   &  \multirow{2}{*}{N/A}      &    \multirow{2}{*}{N/A}    &  \multirow{2}{*}{\$8,000}      &  \multirow{2}{*}{\$40,000} &  \multirow{2}{*}{\$11,200} &  \multirow{2}{*}{N/A} &  \multirow{2}{*}{\textbf{\$59,200}}\\ 

(8\% cluster)  & & & & & & & \\ \hline

Annual (8\% cluster  &  \multirow{2}{*}{N/A}    &   \multirow{2}{*}{N/A}     &  \multirow{2}{*}{\$8,000}     &  \multirow{2}{*}{\$10,000} &  \multirow{2}{*}{\$11,200} &  \multirow{2}{*}{N/A} &  \multirow{2}{*}{\textbf{\$29,200}}\\ 
25\% media) & & & & & & & \\ \hline

\end{tabular}

\caption{Summary of cluster cost estimates of the correlation of a 10-station, dual-polarization, 2 GHz bandwidth per polarization, 1.2 PB data set recorded at 16 Gbps as well as 4.8 GB data set recorded at 64 Gbps with an observation time of $\sim$ 16.7 hours. Both the capital and annual costs are given assuming a five-year replacement for the cluster and disks, linear amortization, disk cost scaling from 16 to 64 Gbps, and prorating of costs assuming 8\% utilization of cluster and 25\% utilization of media.  For simplicity, the five-year cumulative costs of electricity and IT services (0.2 FTE per year or one FTE at an assumed loaded \$200,000) are considered ``capital'' costs. }
\label{clustercost}
\end{table*}
\subsection{Power, cooling, and IT support}
The power consumption of such a cluster is estimated as 30 kW, and assuming an energy cost of \$0.20 per kWh, the power to run the cluster 24/7/365 amounts to \$50,000 per year.  A widely quoted rule-of-thumb for data centers suggests roughly equal power is needed to cool the cluster. With cooling, the total annual electric bill is \$100,000 per year. We assume 0.2 FTE per year of a computer systems technician (``IT'' support) to assist with the assembly and maintenance of the cluster, and estimate \$40,000 for this service.  

\subsection{Amortized total cost}
Under these assumptions, the annualized cost of running the cluster sums to \$250,000 assuming cost of media sufficient for a 16 Gbps experiment. Again, a typical EHT VLBI experiment is run at a bandwidth of 64~Gbps. Although scaling from 16 to 64 Gbps in the cloud quadruples the cost as per Table \ref{totalcost}, the cluster would be more heavily utilized for 64 Gbps, but at no greater marginal cost. This is with the exception of media, for which an annualized cost of \$40,000 to record four times the amount of data must be assumed. This brings the annual cluster cost to \$280,000 for a 64 Gbps experiment. Thus, the annualized cost of the cluster is more than double the $\sim$ \$131,000 cost for a 64 Gbps cloud correlation. 

\subsection{Prorating the cluster for utilization}
The cluster cost estimates assume that it is paid for and run for the full year.  However, it is available to be utilized on other experiments and computation in general when not running VLBI correlations. Making the assumption of perfect utilization at other times --- which assumes excellent operations management --- it is appropriate to prorate the annual cost of the cluster for the time it is actually used during the year.  

We estimate that a 64 Gbps, 10-station, 4.8 PB experiment would require about 11 days of continuous run time for correlation on the hypothetical 1000-core cluster.  Allowing for setup and disk changes, as the cluster only reads and correlates 16 Gbps per run, this might reasonably translate into about a month of wall clock time or about 8\% of the year.  If cluster costs are prorated for 8\% of the year --- though not prorating the costs of media --- the cluster is then the more economical choice at \$59,200.

\subsection{Prorating the media}
We also consider the possibility of prorating media, which can be recycled through multiple observations, as is common in VLBI.  However, the 8\% factor used to prorate the cluster for a month of use is not appropriate in the case of media. The Mark6 disk packs, for instance, are shipped to the sites well ahead of the observation and are conditioned  with read-write cycles there, which is a time-consuming process. After the observation, they are packed and shipped back to the correlator for processing. The data continues to reside on the Mark6 media modules until the correlation job is completed, for which we have assumed one month of clock time.

We estimate two additional months on each side of the observation, to pre-check, pack and ship the disks to the site, including clearing customs; to condition them and then run the observation; and then pack, ship, unpack and stage back at the correlator. In principle, the cluster can be used for other processing jobs during the observation and shipping time, but the disks are not available to recycle into other observations. With the one month of processing in addition to the two months for shipping and staging, we prorate disks to three months, or 25\% of annual. Of course, other observation opportunities are available at exactly the right times for the media to be reused four times a year. Under this assumption, the cluster cost is lower still than the cloud for a 64 Gbps observation. 

\subsection{Cost comparison summary}
Table \ref{clustercost} provides a summary of the three types of scaling relations used for the cluster cost estimates in the prior discussion:  

\begin{itemize}
    \item {\it amortization:} the capital investment considered over an assumed five-year lifetime of the equipment.  
    \item {\it bandwidth scaling:} our benchmarks are for a 16 Gbps recording while a typical EHT wideband experiment is recorded at 64 Gbps.
    \item {\it prorating:} the sharing of capital investment to the benefit of multiple projects.
\end{itemize}

\begin{table*} [htb!]
\centering
\begin{tabular}{|l|c|c|c|c|}
\hline

\multicolumn{1}{|l|}{} & \multirow{2}{*}{Cloud} & \multicolumn{1}{c|}{Cluster} & \multicolumn{1}{c|}{Cluster} & \multicolumn{1}{c|}{Cluster} \\
\multicolumn{1}{|l|}{} & \multicolumn{1}{c|}{} & \multicolumn{1}{c|}{(no prorate)} & \multicolumn{1}{c|}{(8\% cluster prorate)} & \multicolumn{1}{c|}{(8\% cluster, 25\% media prorate)} \\
\hline
\textbf{Total (64 Gbps)}  & \multicolumn{1}{c|}{\textbf{\$130,824}}   & \multicolumn{1}{c|}{\textbf{\$280,000}} & \multicolumn{1}{c|}{\textbf{\$59,200}} & \multicolumn{1}{c|}{\textbf{\$29,200}} \\ 
\hline
\end{tabular}
\caption{Cloud and cluster cost comparison of the correlation of a 10-station, dual-polarization, 4.8 PB dataset recorded at 64 Gbps with 2 GHz bandwidth per polarization with an observation time of $\sim$ 16.7 hours. Three cases are considered for the cluster cost: no prorating, 8\% prorating of the cluster, and 8\% prorating of the cluster as well as 25\% prorating of the media. The prorated cluster costs are lower than the cloud cost. In practice, however, prorating the cluster is almost certainly overly optimistic, especially in the case of media.}
\label{comparecost_table}
\end{table*}

Table \ref{comparecost_table} provides a cloud and cluster cost comparison for the correlation of 4.8 GB data set recorded at 64 Gbps, where the cluster cost is considered under three cases: no prorating, 8\% cluster prorating, and 8\% cluster and 25\% media prorating. We note that the primary benefit of the cloud is not the cost, but rather the ability to deploy much larger computational resources than practically possible on a cluster to run the computation much faster and thereby shorten the time-to-science. This in turn implicitly improves the effective utilization of the entire astronomical instrument, not merely the correlator.  Deploying more cores for a shorter time in the cloud does not significantly affect the cost of the computation. To the contrary, running the computation faster with a greater number of cores reduces the costs of regional storage, which are the dominant cloud costs. However, it is not unreasonable to expect the storage costs on the cloud to decrease with time.

Prorating the cluster and the disk packs is only valid if the rest of  the year the equipment is effectively utilized on other compute jobs (cluster) and observations (media). In practice, the prorating is almost certainly overly optimistic, especially in the case of media, where observations have to become available on precisely a three-month cadence for the 25\% prorating to be achieved. The prorating calculation is useful as a guide to what is possible in principle.  Further, the cluster estimates presented in this subsection are intended only as a rough basis for comparison, with a proper cost comparison being highly assumption dependent, rather complex to put together, and dependent on cost figures which vary greatly over time.

Viewing utilization actually achieved as an economic parameter, there exists a crossover point between the cloud and cluster cost. For low duty cycle wideband observations where substantial cloud resources are strategically deployed, the dominant storage costs for cloud are reduced.  We note that when not utilized, cloud resources are simply released back to the provider, whereas for the cluster, efficient utilization is a significant logistical problem for the operating institution. 

In practice, corporate cloud service providers may provide services at a significant discount, or even gratis, to scientific research groups and non-profits. This practical factor could  dramatically affect the comparative retail costs assumed here, substantially affecting the cost crossover point between the cloud and the cluster.

\section{Summary}
The trend for VLBI observations to record at greater bandwidths has led to increased data volumes. This paper demonstrates the viability of porting wide bandwidth VLBI correlation processing to cloud platforms as a promising alternative to cluster correlation.  

Benchmarks run on 10-station and 20-station synthetic data sets using the \texttt{DiFX-2.5.2} software correlator indicate that full wideband VLBI observations with data volumes exceeding a petabyte can be efficiently processed in elapsed times that are much shorter than the time spent during on-sky observing.  
Similar to some VLBI arrays that transfer data over network and process the correlation in a large computer cluster, the cloud allows the processing of large data sets in parallel. Clusters are often limited by the number of installed compute nodes and playback units as well as the percentage of time allocated for processing different science experiments. Since not every research group has access to a large computer cluster, the advantage of the cloud is its greater accessibility and flexibility that allows researchers to leverage massive compute and storage resources for a short period of time.

Cloud-based computation and data storage enables distributed operation of correlation, allowing international collaborators to efficiently monitor and process VLBI data without the need for physically mounting hard disk media onto localized computing clusters. 

The cloud architecture also ensures built-in upgrades of the rented recording appliances, storage, and compute hardware that can be expected to keep up with industry trends and takes advantage of new commodity equipment as vendors naturally adopt next-generation hardware and software in their cloud platforms. An additional benefit is that all maintenance and operational costs of computing resources are outsourced to the cloud vendor. This study found that the storage of data dominates the overall cost of cloud correlation.

An approximate comparison of cloud to cluster costs is provided. Current cloud service and equipment pricing data is used to compile cost estimates which allow an approximate economic comparison between cloud and cluster processing. Cluster cost is highly dependent on efficient utilization of the physical cluster and associated media. Practically speaking, utilization of cluster and dedicated media inventory is a challenge to optimize, whereas the cloud resources and transfer appliances have costs strictly proportional to time in use.

We intend to continue this research, including looking into increased data rates on the GTAs, and upgrading them to perform reliably at high elevation sites.  A test observation is envisaged as an acid test of the operational framework.  This further research aims to establish whether correlation on the cloud can provide economical extensible path for wideband VLBI arrays to achieve breakthrough science  possible through the efficient analysis of big-data.

\section{Acknowledgments}
This work was supported by grants from the National Science Foundation (AST-1440254, AST-1743747 \& AST-1828513) and the Gordon and Betty Moore Foundation (GBMF-5278). This work was supported in part by the Black Hole Initiative at Harvard University, which is funded by a grant from the John Templeton Foundation. We thank Geoffrey B. Crew, Michael Titus, and  Victor Pankratius for assistance with the \texttt{CorrelX} and \texttt{DiFX-2.5.2} software packages, and Robert W. Wilson for helpful discussions. We gratefully acknowledge support from the Google Cloud Platform to carry out benchmarks in this work. Early developments and benchmarks were also done on the CyVerse Atmosphere Cloud (supported by the National Science Foundation under Award Numbers DBI-0735191 and DBI-1265383) and the Jetstream Cloud \citep{JetStream,JetStream2}. The VLBI software correlation was performed using the \texttt{DiFX-2.5.2} software correlator \citep{Deller_2011}. 

\bibliography{references}

\end{document}